# Glass component induced hysteresis/memory effect in magnetoresistance of ferromagnetic $Pr_{0.9}Sr_{0.1}CoO_{2.99}$

V.P.S. Awana[$], J. Nakamura, M. Karppinen and H. Yamauchi

Materials and Structures Laboratory, Tokyo Institute of Technology, Yokohama 226-8503, Japan

S.K. Malik

Tata Institute of Fundamental Research, Homi Bhabha Road, Mumbai 400005, India

$Pr_{0.9}Sr_{0.1}CoO_{2.99}$ sample exhibits magnetoresistivity (MR) of up to 40 % at 5 K with a strong hysteresis/memory effect. Magnetisation measurements on $Pr_{0.9}Sr_{0.1}CoO_{2.99}$ in an applied field of 100 Oe show that, as temperature decreases, the zero-field-cooled (ZFC) and field-cooled (FC) magnetisation curves branch clearly at 50 K, and a cusp appears in the ZFC branch at $T_{cusp}$ ~20 K. Magnetisation measurements in various fields between 100 and 10,000 Oe show that both the ZFC-FC branching temperature and, $T_{cusp}$, decrease with increasing field. The magnetization-field isotherms at 5 and 10 K show hysteresis loops typical of ferromagnets. No appreciable MR is seen in this compound at 50 K, *i.e.* at a temperature close to ZFC-FC branching temperature. At 20 K, negative MR of above 16% is observed without any hysteresis effect. We believe that the appearance of a ferromagnetic component at 5 K and 10 K (i.e. at temperatures below $T_{cusp}$) within the spin glass state of Co spins is responsible for both large MR and the prominent hysteresis/memory effect in MR.

[$] awana1@rlem.titech.ac.jp : Fax No. 0081-45-924 5339

## I. Introduction

Observations of magnetoresistance (MR) or colossal magnetoresistance (CMR) in compounds with formulae, $R_{1-x}A_x MnO_{3-\delta}$ (where $R$ = La, Pr, Nd, Sm, etc., and $A$ = Ca, Sr, Ba, Pb, etc.) have attracted a lot of attention [1,2]. The Mn-based perovskites, exhibiting MR/CMR have been extensively studied in the last few years [1-3]. More recently, the focus has shifted to some other extended-valency transition metal compounds, such as Co- and Fe-based perovskite compounds [4-6].

In the present contribution, we focus our attention on the single perovskite Co-based oxide, $Pr_{1-x}Sr_x CoO_{3-\delta}$. The $(Pr/La)_{1-x}Sr_x CoO_{3-\delta}$ compounds have been widely investigated for possible uses in electrochemical devices [7]. In these compounds, changes in x and/or $\delta$ bring about a sharp metal-insulator transition [8]. Besides practical uses of $La_{1-x}Sr_x CoO_{3-\delta}$ for electrochemical devices, it was recently recognized that these compounds exhibit MR of up to 30 % for x < 0.1 [9]. Magnetization studies and pertaining magnetic structure of various similar cobaltates have been investigated in the past [10-14]. Also attempts have been made to relate the spin glass behavior of Co spins and the ensuing MR properties [11,13,14-16]. One of the interesting observations is that of the memory effect or the hysteresis [15] in MR of some cobaltates - an effect which is yet to be understood. Keeping this in view, we have studied $Pr_{0.9}Sr_{0.1}CoO_{2.99}$ for its magnetic and transport properties. This compound is shown to exhibit MR of up to 40 % at 5 K, with a strong memory/hysteresis effect. Magnetization of this compound shows the presence of a weak ferromagnetic (FM) component along with glassy Co spins. However, the MR is observed in the whole ferromagnetic region, the memory/hysteresis effect in MR is seen only in the region where both ferromagnetism and the spin-glass state co-exist. Our results demonstrate that the memory effect in MR of this compound basically arises from slow relaxation of the remnant field due to the presence of some spin glass component.

## II. Experimental details

Samples of $Pr_{0.9}Sr_{0.1}CoO_{3-\delta}$ were synthesized through a solid-state reaction route from $Pr_6O_{11}$, $SrO_2$ and $Co_2O_3$. Calcinations were carried out on mixed powders at 1000 $^0$C, 1025 $^0$C, 1050 $^0$C and 1075 $^0$C each for 24 hours with intermediate grindings. The pressed bar-shape pellets were annealed in a flow of oxygen at 1100 $^0$C for 40 hours and subsequently cooled slowly, over a span of 20 hours, to room temperature. X-ray diffraction (XRD) data were obtained at room temperature (MAC Science: MXP18VAHF[22]; Cu$K_\alpha$ radiation). Magnetization measurements were carried out on a SQUID magnetometer (Quantum Design: MPMS-5S). Resistivity measurements in applied magnetic fields between 0 and 7 T, and in the temperature range of 5 - 300 K, were performed using a four point method (Quantum Design: PPMS). The oxygen content of the sample was established at $\delta = 0.01$ by means of wet-chemical cerimetric titration [17].

### III. Results and discussion

Refinement of the X-ray diffraction data for $Pr_{0.9}Sr_{0.1}CoO_{2.99}$ revealed that the compound crystallizes in an orthorhombic $GdFeO_3$-type structure (space group, *Pnma*). The lattice parameters determined are: $a = 5.365(2)$ Å, $b = 7.606(1)$ Å, and $c = 5.418(3)$ Å, which are in agreement with those reported earlier [12] for fully oxygen annealed $Pr_{0.9}Sr_{0.1}CoO_{2.99}$ sample.

Figure 1 shows the DC susceptibility *vs.* temperature *($c$-T)* behavior for $Pr_{0.9}Sr_{0.1}CoO_{2.99}$ in an applied field of 100 Oe. The zero-field-cooled (ZFC) and field-cooled (FC) curves branch clearly at 50 K. Further, a cusp in the susceptibility appears in the ZFC curve only at $T_{cusp}$ ~20 K. Such a cusp is indicative of either a spin glass (SG) state or an antiferromagnetically (AFM) ordered state of the Co spins. As temperature decreases, the FC measured magnetization increases and later saturates (at low temperatures), without forming a cusp. The present magnetization results are in agreement with those reported earlier on $Pr_{1-x}Sr_xCoO_{3-\delta}$ samples with x < 0.1 [18]. To understand the nature of magnetic ordering at the ZFC-FC branching temperature and at $T_{cusp}$ we carried out magnetization measurements under various magnetic fields of 500, 1,000 and 10,000 Oe. The data are plotted in the inset of Fig. 1. Both the ZFC-FC branching temperature (upward arrows) and $T_{cusp}$ (downward arrows), are seen to decrease with increasing field strength. These observations exclude the possibility of AFM ordering of the Co spins [19]. The decrease in the ZFC-FC branching temperature with increasing field strength is indicative of ferro- and/or ferri-magnetic ordering of the Co spins. Hence it is likely that the cusp in the ZFC branch stems from a spin-glass state of the Co spins. In short, with decreasing temperature, the Co spins order ferro-/ferri-magnetically at the ZFC-FC branching temperature (~50 K) and then some spin-glass component develops below $T_{cusp}$.

To ascertain the presence of ferro-/ferri-magnetic component, we plot the magnetization ($M$) vs. applied field ($H$) for $Pr_{0.9}Sr_{0.1}CoO_{2.99}$ at 5 K and 10 K (Fig. 2). Hysteresis loops are clearly seen at both these temperatures with coercive fields of around 0.3 T and 0.2 T at 5 and 10 K, respectively. The hysteresis loops, are indicative of ferro-/ferri-magnetic spin contributions. As the FC part (Fig. 1) of magnetization shows neither typical down-turn nor subsequent saturation in magnetization to the lowest measured temperature, the possibility of ferrimagnetic component is excluded. In fact, saturating FC part at low temperatures (below 15 K) indicates a ferromagnetic component. Thus we conclude that the hysteresis loops observed at 5 and 10 K are due to the presence of ferromagnetic Co spin component in the compound. To further elucidate this point, we show in the inset of Fig. 2 isothermal magnetization ($M$) vs. field ($H$) curves measured at 25, 50, 100 and 150 K. Above the clear ZFC-FC branching temperature of ~50 K (Fig.1), the compound is paramagnetic and hence the $M$ vs $H$ relation is linear. Below 50 K, the ferromagnetic component due to the Co spins starts developing and then non-linearity appears in the $M$ vs. $H$ curves. The isothermal magnetization ($M$) as a function of magnetic field ($H$) below the ordering temperature may be written as:

$$M(H) = \chi H + \sigma_s (H), \qquad (1)$$

where $\chi H$ is the linear contribution from paramagnetic Co and Pr spins and $\sigma_s$ represents the ferromagnetic component due to the Co spins, which appears only below the clear ZFC-FC branching temperature of 50 K.

The Co spins in $Pr_{0.9}Sr_{0.1}CoO_{2.99}$ are paramagnetic at room temperature, and develop a non-linear contribution to isothermal magnetization below around 50 K due to ferromagnetism. At still lower temperatures, spin-glass component develops below $T_{cusp}$, which also contributes to the non-linearity in M-H curves. The magnetic structure of $Pr_{0.9}Sr_{0.1}CoO_{2.99}$ at 5 K may be viewed in terms of a spin-glass component within the ferromagnetically ordered Co spins, similar to that reported for $La_{0.9}Sr_{0.1}CoO_3$ [10]. Such a spin-glass structure was also confirmed for $Pr_{1-x}Sr_xCoO_{3-\delta}$ from the relaxation process of magnetization as a function of time [18,20].

Figure 3 shows the resistivity ($\rho$) vs. Temperature plot for $Pr_{0.9}Sr_{0.1}CoO_{2.99}$: both heating (5 → 300 K) and cooling (300 → 5 K) data are shown. There are essentially no differences between the cooling and the heating data in the temperature range of 20 to 300 K. Below 20 K, the $\rho$ values are slightly less in cooling than in heating. The resistivity behavior of the compound can be fitted to a thermal activation process:

$$\rho = \rho_0 \exp(E_a/k_B T), \qquad (2)$$

where $\rho_0$ is a constant, $E_a$ the activation energy and $k_B$ the Boltzman constant. Inset in Fig.3 shows the $\ln(\rho)$ vs. $1/T$ plot, which is linear between 50 and 300 K but changes its slope below this temperature range. From the $\ln(\rho)$ vs. $1/T$ data in the temperature range of 50 to 300 K, a value of 1.08 eV is obtained for $E_a$. The sharp change in the slope of the $\ln(\rho)$ vs. $1/T$ plot below 50 K most likely indicates a change in the conduction process. In fact the $\rho(T)$ behaviour below 20 K can be fitted to a 3D-VRH (variable range hopping) conduction process. Interestingly this temperature of 20 K is close to $T_{cusp}$, below which a spin-glass component appears in the magnetization. Thus $Pr_{0.9}Sr_{0.1}CoO_{2.99}$ is basically a doped semiconductor which follows a VRH process in electrical conduction at low temperatures below 20 K. Recently we reported a change in electrical conduction with temperature in $R_{1/3}Sr_{2/3}FeO_{3-\delta}$ compounds also with both $R$ = La and Pr [21].

Figure 4 shows the magnetoresistance of $Pr_{0.9}Sr_{0.1}CoO_{2.99}$ in various applied fields at temperatures of 5 K, 20 K and 50 K. Here the MR is calculated using the following formula:

$$MR\% = [(\rho_H - \rho_0)/\rho_H] \times 100, \qquad (3)$$

where $\rho_H$ is the resistivity in the magnetic field $H$ and $\rho_0$ is the corresponding value in zero external magnetic field. No appreciable MR is observed in this compound at 50 K. Negative MR of up to 16 % is observed under applied magnetic field of 7 T, at 20 K. At 5

K, with increasing field strength the resistivity increases to a maximum ($\rho_{peak}$) at a field lower than 1 T and later decreases. The MR values in Fig. 4 are normalized to the $\rho_{peak}$ value. The compound shows negative MR of up to 40 % under 7 T applied field at 5 K. Unlike at 20 K, one clearly sees a memory/hysteresis effect in MR at 5 K. That is, the MR values depend not only on the applied field strength but also on the route for increasing/decreasing the field strength.

The magnetization and magnetotransport results for $Pr_{0.9}Sr_{0.1}CoO_{2.99}$ may be summarized as follows:

1. The Co spins in $Pr_{0.9}Sr_{0.1}CoO_{2.99}$ are paramagnetic at room temperature but order magnetically at 50K. A non-linear contribution to the isothermal magnetization develops below *50 K* due to ferromagnetic interactions between Co spins and later due to a spin-glass component below 20 K.
2. The $Pr_{0.9}Sr_{0.1}CoO_{2.99}$ oxide system is basically a moderately doped semiconductor, following a VRH process for electrical conduction at low temperatures and exhibiting thermal hysteresis in the $\rho(T)$ behaviour below 20 K.
3. The $Pr_{0.9}Sr_{0.1}CoO_{2.99}$ oxide system shows MR of up to 40 % in 7 T applied field at 5 K with a prominent hysteresis/memory effect.

Presently various MR/CMR materials are being investigated [1-6]. In the case of widely studied $R_{1-x}A_x MnO_{3-\delta}$ (where $R$ = La, Pr, Nd, Sm, etc., and $A$ = Ca, Sr, Ba, Pb, etc.) compounds, the mechanism of MR is more or less understood. The suppression of the charge-ordered (CO) state or the structural change under high magnetic fields and the resultant sharp decrease in resistance are essential in giving rise to CMR [3,22]. Recently, magnetic phase separation was recognised to play an important role in MR [23,24]. These factors are favourable for the appearance of MR, but do not necessarily guarantee MR of reasonable magnitudes. For example, for $R_{1/3}Sr_{2/3}FeO_{3-\delta}$ with $R$ = La and Pr, a field of 7 T is not sufficient to alter the CO state and hence only weak MR of up to few percent was observed [21]. However, in the case of $Nd_{0.5}Sr_{0.5}MnO_3$, this field is sufficient for suppressing the CO state to turn the compound from an insulating state to a nearly metallic state, giving rise to a huge MR [25]. For the $Pr_{0.9}Sr_{0.1}CoO_{2.99}$, CO state has not been observed in the temperature range studied [4,9,12,18,20]. The present magnetization data on $Pr_{0.9}Sr_{0.1}CoO_{2.99}$ show the existence of a spin-glass component in ferromagnetically ordered state of the Co spins. The spin-glass state is responsible for the thermal history presently observed below 20 K in $\rho(T)$ data (Fig. 3). Also in this temperature range, the electrical conduction switches from a moderately doped semiconducting process to a 3D-VRH process. It seems that the 3D-VRH process gets frustrated due to de-localization of the hopping carriers under magnetic fields and hence MR appears. Worth mentioning is the fact that no appreciable MR is seen at 50 K, i.e. well above the 3D-VRH process and the spin-glass state. At 20 K, the MR is seen up to 16 %, but without any real history effect. Interestingly this temperature (20 K) is close to the $T_{cusp}$, where the spin-glass component starts appearing in the system. Most likely, the coexistence of a spin-glass state and a ferromagnetically ordered state of the Co spins at 5 K results in slowing down of the relaxation process of the internal magnetic fields. The ensuing slow relaxation of the remnant field increases the effective value of the externally applied field inside the compound. This results in a larger MR due to the higher effective field (= applied field + remnant of the last applied field in the sequence). For example, in an applied field of 3 T, the net applied field is 3 T + remnant of 4 T when decreasing field and 3 T + remnant of 2 T when increasing field. The higher value of the net applied field, when decreasing the field, gives rise to the higher negative MR than when increasing field at the same applied field. This causes the MR memory/hysteresis effect, which basically arises from slow relaxation of the remnant field due to the presence of some spin-glass component. This gets credence from the fact that though MR of up to 16 % is observed at 20 K, no hysteresis is seen in the same due to lack of spin glass component in the compound at this temperature.

## IV. Conclusions

The oxide, $Pr_{0.9}Sr_{0.1}CoO_{2.99}$, is basically a doped semiconductor, following below 20 K a 3D-VRH conduction process with thermal hysteresis in the $\rho(T)$ behaviour. The Co spins in $Pr_{0.9}Sr_{0.1}CoO_{2.99}$ are paramagnetic at room temperature, and develop ferromagnetic interactions and some spin-glass component below 50 and 20 K, respectively. This compound shows MR of up to 40 % at 5 K under 7 Tesla and exhibits a memory effect. The coexistence of spin-glass and ferromagnetically ordered Co spins at low temperatures is responsible for both MR and the prominent memory/hysteresis effect. The MR memory effect of $Pr_{0.9}Sr_{0.1}CoO_{2.99}$ may qualify the material for practical uses.

**Acknowledgements**

Ms. K. Salömaki is thanked for her help in cerimetric titrations. The present work has been supported by a Grant-in-Aid for Scientific Research (Grant No. 11305002) from the Ministry of Education, Science and Culture of Japan, by an International Collaborative Research Project Grant-2000 of Materials and Structures Laboratory, Tokyo Institute of Technology, and by the Academy of Finland (Decision No. 46039). V.P.S.A. acknowledges the Japanese Society for Promotion of Sciences for the Foreigners Post Doctoral Fellowship with ID P-00128.

**Figure captions**

Figure 1. Susceptibility *vs*. temperature plot for $Pr_{0.9}Sr_{0.1}CoO_{2.99}$ in an applied field of 100 Oe. Both ZFC and FC parts are shown. Inset shows the same in various applied fields of 500, 1000 and 10000 Oe.

Figure 2. *M vs. H* hysteresis loops for $Pr_{0.9}Sr_{0.1}CoO_{2.99}$ at 5 and 10 K. Inset shows the isothermal magnetization *vs*. applied field for the same at 25, 50, 100 and 150 K.

Figure 3. Resistivity *vs*. temperature plot for $Pr_{0.9}Sr_{0.1}CoO_{2.99}$. Iinset shows $\ln\rho$ *vs*. $1/T$ plot for the same.

Figure 4. Magnetoresistance *vs*. applied field plots for $Pr_{0.9}Sr_{0.1}CoO_{2.99}$ at 5, 20 and 50 K. The hysteresis in MR is seen at 5 K.

*Fig. 1. Awana etal. Letter to Editor JMMM.*

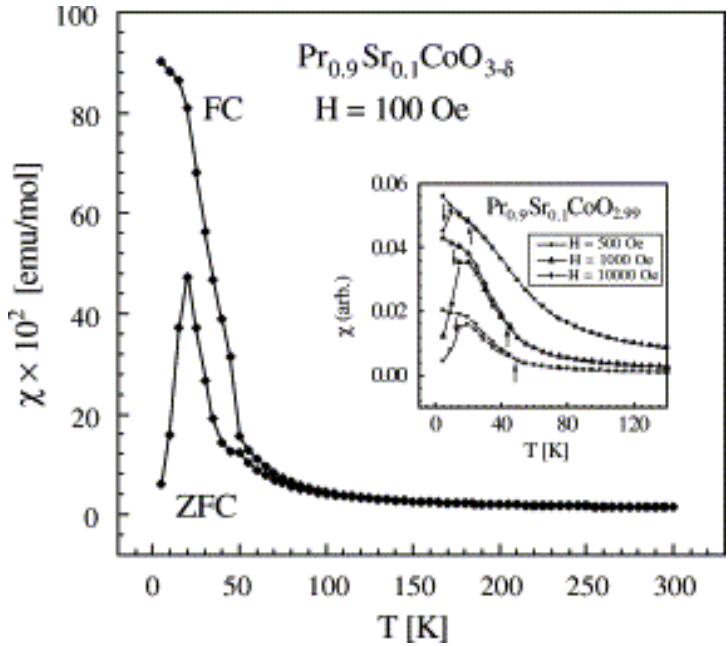

*Fig.2 Awana etal. Letter to The editor JMMM*

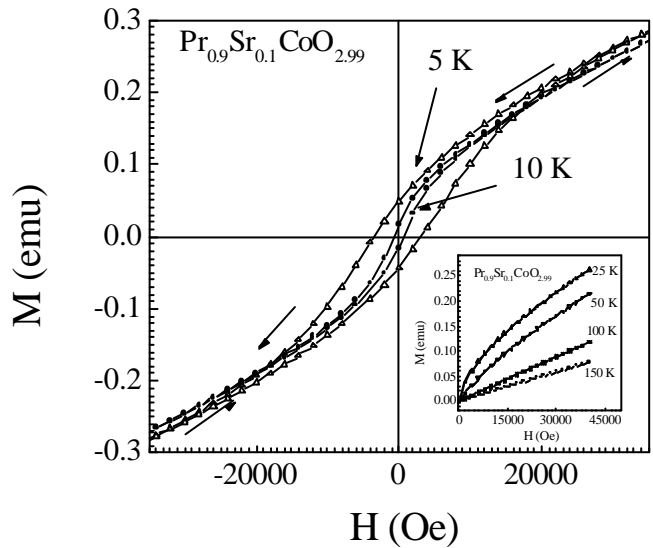

*Fig.3 Awana etal. Letter to the Editor JMMM*

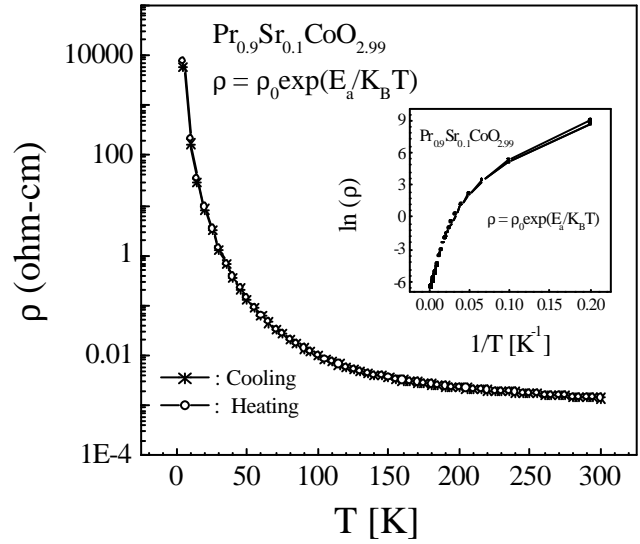

*Fig. 4 Awana etal. Letter to The editor JMMM*

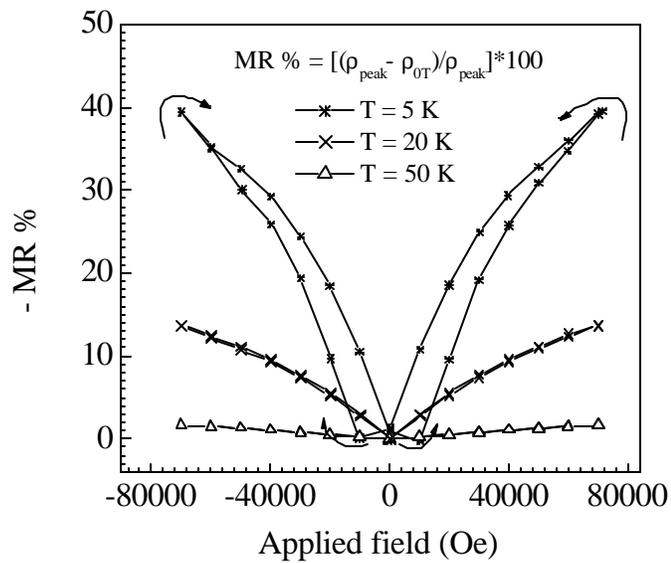